\documentstyle[12pt]{article}
\begin{document}
\begin{flushright}
{\small Submitted for publication in the Proceedings of the 1998 \\
Ringberg symposium ``Fields, Particles and Gravitation" \\
(on the occassion of the W. Zimmermann 70th birthday)}
\end{flushright}
\vspace{3mm}

\begin{center}
{\large\bf The Bogoluibov Renormalization Group \\ in Theoretical
and Mathematical Physics}  \vspace{3mm}

D.V.~Shirkov \\
\vspace{2mm}

  {\it N.N.Bogoliubov Laboratory of Theoretical Physics, \\
JINR, Dubna, Russia; \ \ shirkovd@thsun1.jinr.ru}
\end{center}
\vspace{2mm}

\abstract{This text follows the line of a talk on Ringberg symposium
dedicated to Wolfhart Zimmermann 70th birthday. The historical overview
(Part 1) partially overlaps with corresponding text of my previous
commemorative paper -- see Ref. \cite{umn94} in the list. At the same time
second part includes some recent results in QFT (Sect. 2.1) and summarize
(Sect. \ref{mf}) an impressive progress of the ``QFT renormalization group"
application in mathematical physics.}
\tableofcontents

\newpage   % \pagenumbering{arabic}

\section{Early History of~the ~RG~in ~the~QFT}

\subsection{The birth of Bogoliubov's renormalization group.}
      In the spring of 1955 a small conference on ``Quantum Electrodynamics
and Elementary Particle Theory'' was organized in
Moscow. It took place at the Lebedev Institute in the first half of April.
Among the participants there were several foreigners, including Hu Ning
and Gunnar K\"all\'en.
    Landau's survey lecture ``Fundamental Problems in QFT'', in which the
issue of ultraviolet (UV) behaviour in the QFT was discussed, constituted the
central event of the conference.  Not long before, the problem of
short-distance behaviour in QED was advanced substantially in a series of
articles \cite{lakh} by Landau, Abrikosov, and Khalatnikov. They succeeded in
constructing a closed approximation of the Schwinger--Dyson equations, which
 admitted an explicit solution in the massless limit and, in modern language,
it resulted in the summation of the leading UV logarithms. \par

    The most remarkable fact was that this solution turned out to be
self--contradictory from the physical point of view because it contained a
``ghost  pole'' in the renormalized amplitude of the photon propagator or, in
terms of bare notions, the difficulty of ``zero physical charge''. \par

    At that time our meetings with Nicolai Nicolaevich Bogoliubov (N.N. in
what follows) were regular and intensive because we were tightly involved in
the writing of final text\footnote{ Just at that time the first draft of a
central part of the book has been published \cite{ufn55} in the form of two
\par extensive papers.} of our big book. N.N. was very interested in the
results of Landau's group and proposed me to consider the general problem of
evaluating their reliability by constructing, e.g., the second approximation
(including {\it next-to-leading UV logs}) to the Schwinger--Dyson equations,
to verify the stability of the UV asymptotics and the very existence of a
ghost pole. \par

  Shortly after the meeting at the Lebedev Institute, Alesha Abrikosov told
me about Gell--Mann and Low's article\cite{gml} which had just appeared.
The same physical problem was treated in this paper, but, as he put it,
it was hard to understand and to combine it with the results obtained by the
Landau group. \par
   I looked through the article and presented N.N. with a brief report on the
methods and results, which included some general assertions on the scaling
properties of the electron charge distribution at short distances and rather
cumbersome functional equations -- see, below, Section ~\ref{1-3}. \par

   N.N. immediate comment was that Gell--Mann and Low's approach is very
important: it is closely related to the {\it la groupe de normalisation}
discovered a couple of years earlier by Stueckelberg and Petermann \cite{stp}
in the course of discussing the structure of the finite arbitrariness in the
scattering matrix elements arising upon removal of the divergences. This
group is an example of the continuous groups studied by Sophus Lie. This
implied that functional group equations similar to those of paper \cite{gml}
should take place not only in the UV limit but also in the general case as
well. \par

 Within the next few days I succeeded in recasting Dyson's finite
transformations and obtaining the desired functional equations for the QED
propagator amplitudes, which have group properties, as well as the group
differential equations, that is, the Sophus Lie equations of the
renormalization group (RG). Each of these resulting equations --- see, below
Eqs.(\ref{feqs} --- contained a specific object, the product of the squared
electron charge $\alpha=e^2$ and the transverse photon propagator amplitude
$d(Q^2)$. We named this product, $e^2(Q^2)=e^2 d(Q^2)$, the {\it invariant
charge}. From the physical point of view this function is an analogue of the
so--called {\it effective charge} of an electron, first discussed by Dirac in
1933~\cite{dirac}, which describes the effect of the electron charge
screening due to quantum vacuum polarization. Also, the term
``renormalization group" was first introduced in our Doklady Akademii Nauk
SSSR publication \cite{bs-55a} in 1955 (and in the English language paper
\cite{nc-56}).

  At the above--mentioned Lebedev meeting Gunnar K\"all\'en presented a paper
written with Pauli on the so--called ``Lee model'', the exact solution of
which contained a {\it ghost  pole} (which, in contrast to the physical one
corresponding to a bound state, had negative residue) in the nucleon
propagator. K\"all\'en--Pauli's analysis led to the conclusion that the Lee
model is physically void. \par

 In view of the argument on the presence of a similar pole in the QED photon
propagator (which follows from the abovementioned solution of Landau's group
as well as from an independent analysis by Fradkin \cite{efim}) obtained in
Moscow, K\"all\'en's report resulted in a heated discussion on the possible
inconsistency of QED. In the discussion K\"all\'en argued that no rigorous
conclusion about the properties of sum of an infinite nonconvergent series
can be drawn from the analysis of a finite number of terms. \par

  Nevertheless, before long a publication by Landau and Pomeranchuk (see,
e.g., the review paper\cite{dau-bohr}) appeared arguing that not only QED
but also local QFT were self--contradictory. \par
  Without going into details, remind that our analysis of this problem
carried out \cite{bsh56} with the aid of the RG formalism just appeared led
to the conclusion that such a claim cannot have the status of a {\it rigorous
result, independent of perturbation theory}. \par

\subsection{Renormalization and renormalization invariance.\label{1-2}}

    As is known, the regular formalism for eliminating ultraviolet
divergences in quantum field theory (QFT) was developed on the basis of
covariant perturbation theory in the late 40s. This breakthrough is connected
with the names of Tomonaga, Feynman, Schwinger and some others. In
particular, Dyson and Abdus Salam carried out the general analysis of the
structure of divergences in arbitrarily high orders of perturbation theory.
Nevertheless, a number of subtle questions concerning so-called overlapping
divergences remained unclear. \par
   An important contribution in this direction based on a thorough analysis
of the mathematical nature of UV divergences was made by Bogoliubov. This was
achieved on the basis of a branch of mathematics which was new at that time,
namely, the Sobolev--Schwartz {\it theory of distributions}. The point is
that propagators in local QFT are distributions (similar to the Dirac
delta--function) and their products appearing in the coefficients of the
scattering matrix expansion require supplementary definition in the case when
 their arguments coincide and lie on the light cone. In view of this the UV
divergences reflect the ambiguity in the definition of these products.\par In
  the mid 50ies on the basis of this approach Bogoliubov and his disciples
developed a technique of supplementing the definition of the products of
singular Stueckelberg--Feynman propagators \cite{ufn55} and proved a
theorem~\cite{paras} on the finiteness and uniqueness (for renormalizable
theories) of the scattering matrix in any order of perturbation theory. The
prescription part of this theorem, namely, {\it Bogoliubov's R-operation},
still remains a practical means of obtaining finite and unique results in
perturbative calculations in QFT. \par

   The Bogoliubov algorithm works, essentially, as follows: \par
-- To remove the UV divergences of one-loop diagrams, instead of introducing
some regularization, for example, the momentum cutoff, and handling (quasi)
infinite counterterms, it suffices to complete the definition of
divergent Feynman integral by subtracting from it certain polynomial in the
external momenta which in the simplest case is reduced to the first few terms
of the Taylor series of the integral. \par
-- For multi-loop diagrams (including ones with overlapping divergencies) one
 should first subtract all divergent subdiagrams in a hierarchical
 order regulated by the $\,R$--operator. \par
 The uniqueness of computational results is ensured by special conditions
 imposed on them. These conditions contain specific degrees of freedom
(related to different renormalization schemes and momentum scales) that
can be used to establish the relationships between the Lagrangian parameters
(masses, coupling constants) and the corresponding physical quantities. The
fact that physical predictions are independent of the arbitrariness in the
renormalization conditions, that is, they are {\it renorm--invariant},
constitutes the conceptual foundation of the renormalization group. \par

 An attractive feature of this approach is that it is free from any auxiliary
nonphysical attributes such as bare masses, bare coupling constants, and
regularization parameters which turn out to be unnecessary in computations
employing Bogoliubov's approach. As a whole, this method can be regarded as
{\it renormalization without regularization and counterterms}.

\subsection{The discovery of the renormalization group.\label{1-3}}
 The renormalization group was discovered by Stueckelberg and Petermann
\cite{stp} in 1952-1953 as a group of infinitesimal transformations related
to a finite arbitrariness arising in the elements of the scattering
$S$-matrix upon elimination of the UV divergences. This arbitrariness can be
fixed by means of certain parameters $c_i$:
\begin{quote}
{\sl ``... we must expect that a group of infinitesimal operators
 $\mbox{\bf P}_i=(\partial/\partial c_i)_{c=0}$,
exists, satisfying
$$
\mbox{\bf P}_iS=h_i(m,e)\partial S(m,e,...)/\partial e~, $$
admitting thus a renormalization of $e$."}
\end{quote}
\noindent These authors introduced the {\it normalization group} generated
(as a Lie group) by the infinitesimal operators $\mbox{\bf P}_i$ connected
with renormalization of the coupling constant $e$. \par
    In the following year, on the basis of Dyson's transformations written in
the regularized form, Gell-Mann and Low~\cite{gml} derived functional
equations for QED propagators in the UV limit. For example, for the
renormalized transverse part $d$ of the photon propagator they obtained an
equation of the form
\begin{equation}
\label{1}
d\left(\frac{k^2}{\lambda^2},e_2^2\right)=
\frac{d_C(k^2/m^2,e_1^2)}{d_C(\lambda^2/m^2,e_1^2)}~,~~
e_2^2=e_1^2d_C(\lambda^2/m^2,e_1^2)~,
\end{equation}
where $\lambda$  is the cutoff momentum and  $e_2$ is the physical
electron charge. The appendix to this article contains the general
solution (obtained by T.D.Lee) of this functional equation for the
photon amplitude $d(x,e^2)$ written in two equivalent forms:
\begin{equation}\label{2}
e^2d\left(x,e^2\right)=F\left(xF^{-1}\left(e^2\right)\right)
\;,\;\;\;\;
\ln x=\int\limits_{e^2}^{e^2d}\frac{{\rm d}y}{\psi(y)}~,
\end{equation}
with
$$
\psi(e^2)=\frac{\partial(e^2d)}{\partial\ln x}~~~\mbox{at}~~~x=1~. $$
A qualitative analysis of the behaviour of the electromagnetic
interaction at small distances was carried out with the aid of
(\ref{2}). Two possibilities, namely, infinite and finite charge
renormalizations were pointed out:
 \begin{quote}
{\sl
\noindent Our conclusion is that the {\bf shape} of the charge distribution
surrounding a test charge in the vacuum does not, at small distances, depend
on the coupling constant except through the scale factor. The behavior of the
propagator functions for large momenta is related to the magnitude of the
renormalization constants in the theory. Thus it is shown that the
unrenormalized coupling constant $e_0^2/4\pi\hbar c$, which appears in
perturbation theory as a power series in the renormalized coupling constant
$e_1^2/4\pi\hbar c$ with divergent coefficients, many behave either in two
ways:

It may really be infinite as perturbation theory indicates;

It may be a finite number independent of $e_1^2/4\pi\hbar c$. }
 \end{quote}

  Note, that the latter possibility corresponds to the case when $\psi$
vanishes at a finite point:  $\psi(\alpha_\infty)=0$.  Here, $\alpha_\infty$
is known now as a fixed point of the renormalization group transformations.

The paper~\cite{gml} paid no attention to the group character of the analysis
and the results obtained there.  The authors failed to establish a connection
between their results and the standard perturbation theory and did not
discuss the possibility that a ghost pole might exist.

  The final step was taken by Bogoliubov and Shirkov~\cite{bs-55a,bs-55b} --
see also the survey~\cite{nc-56} published in English in 1956. Using the
group properties of finite Dyson transformations for the coupling constant
and the fields, these authors derived functional group equations for the
propagators and vertices in QED in the general case (that is, with the
electron mass taken into account). For example, the equation for the
transverse amplitude of the photon propagator and electron propagator
amplitude were obtained in the form
\begin{equation} \label{feqs}
d(x,y;e^2)=d(t,y;e^2)d\left(\frac{x}{t},\frac{y}{t};e^2d(t,y;e^2)\right),\;
s(x,y;e^2)=s(t,y;e^2) s\left(\frac{x}{t}, \frac{y}{t}; e^2d(t,y;e^2)\right)
\end{equation}
in which the dependence not only on momentum transfer $x=k^2/\mu^2$
(where $\mu$ is a certain normalizing scale factor), but also on the mass
variable $y=m^2/\mu^2$ is taken into account.  \par
 As can be seen, the product  $e^2d\,$ of electron charge squared and photon
propagator amplitude enters in both functional equations. This product is
invariant with respect to Dyson transfermation. We called this function --
{\it invariant charge}. \par
  In the modern notation, the first equation (which in the massless case
$\,y=0\,$ is equivalent to (\ref{1})) is an equation for the invariant charge
(now widely known as an effective or running coupling)
$\bar\alpha =\alpha d(x,y;\alpha=e^2)$:
\begin{equation}
\label{3}      \bar\alpha(x,y;\alpha)=
\bar\alpha\left(x/t, y/t; \bar\alpha(t,y;\alpha)\right)~.
\end{equation}
   Let us emphasize that, unlike in the Ref.\cite{gml} approach, in our case
there are no simplifications due to the massless nature of the UV asymptotics.
Here the homogeneity of the transfer momentum scale is violated explicitly by the
mass $m$. Nevertheless, the symmetry (even though a bit more complex one)
underlying the renormalization group, as before, can be stated as an {\it
exact symmetry} of the solutions of the quantum field problem -- see eq.
(\ref{10}) below. This is what we mean when using the term {\it Bogoliubov's
renormalization group} or {\it renorm-group} for short.

The differential group equations (DGEs) for $\bar\alpha$ and for the electron
propagator:
\begin{equation}\label{4}
\frac{\partial\bar\alpha(x,y;\alpha)}{\partial\ln x}=
\beta\left(\frac{y}{x},\bar\alpha(x,y;\alpha)\right)\,\,;\,\,\,
\frac{\partial s(x,y;\alpha)}{\partial\ln x}=
\gamma\left(\frac{y}{x},\bar\alpha(x,y;\alpha)\right)s(x,y;\alpha)~,
\end{equation}
with
\begin{equation} \label{6}
\beta(y,\alpha)=\frac{\partial\bar\alpha(\xi,y;\alpha)}{\partial\xi}~,~
~~~\gamma(y,\alpha)=\frac{\partial s(\xi,y;\alpha)}{\partial\xi}~~~~
\mbox{at}~~\xi=1~.
\end{equation}
were first derived in \cite{bs-55a} by differentiating the functional
equations. In this way an explicit realization of the DGEs mentioned in the
citation from~\cite{stp} was obtained. These results established a conceptual
link with the Stueckelberg--Petermann and Gell-Mann -- Low approaches.

\subsection{Creation of the RG method\label{1-4}}

Another important achievement of paper \cite{bs-55a} consisted in formulating a
simple algorithm for improving an approximate perturbative solution by
combining it with the Lie differential equations (modern notation is
used in this quotation from~\cite{bs-55a}):

\begin{quote}
{\sl Formulae (\ref{4}) show that to obtain expressions for $\bar\alpha$ and
$s$ valid for all values of their arguments one has only to define
$\,\bar\alpha(\xi,y,\alpha)\,$ and $s(\xi,y,\alpha)\,$ in the vicinity of
$\xi=1$. This can be done by means of the usual perturbation theory.}
\end{quote}

 In our adjacent publication~\cite{bs-55b} this algorithm was effectively
used to analyse the UV and infrared (IR) asymptotic behaviour in QED. The
one-loop and two-loop UV asymptotics
\begin{equation} \label{a1rg}
\bar\alpha^{(1)}_{RG}(x;\alpha)\equiv \bar\alpha^{(1)}_{RG}(x,0,\alpha)=
\frac{\alpha}{1-\frac{\alpha}{3\pi}\cdot\ln x}\,\,,  \end{equation}
\begin{equation}   \label{a2rg}
\bar\alpha^{(2)}_{RG}(x;\alpha)= \frac{\alpha}{1-\frac{\alpha}{3\pi}\ln x
+\frac{3\alpha}{4\pi}\ln(1-\frac{\alpha}{3\pi}\ln x)}
\end{equation}
of the photon propagator as well as the IR asymptotics
$$
s(x,y;\alpha)\approx (x/y-1)^{-3\alpha/2\pi}=(p^2/m^2-1)^{-3\alpha/2\pi}~$$
of the electron propagator in transverse gauge were obtained. At that time
these expressions had already been known only at the one--loop level. It
should be noted that in the mid 50s the problem of the UV behaviour in local
QFT was quite urgent.  As it has been mentioned already a substantial
progress in the analysis of QED at small distances was made by Landau and his
collaborators~\cite{lakh}.  However, Landau's approach did not provide a
prescription for constructing subsequent approximations.

 An answer to this question was found only within the new renorm--group
method. The simplest UV asymptotics of QED propagators obtained in our
paper~\cite{bs-55b}, for example, expression (\ref{a1rg}), agreed precisely
with the results of Landau's group.

Within the RG approach these results can be obtained in just a few lines of
argumentation. To this end, the massless one-loop approximation
$$
\bar\alpha^{(1)}_{PTh}(x;\alpha)=\alpha+
\frac{\alpha^2}{3\pi}\ell+...~~,~~~~\ell=\ln x $$
of perturbation theory should be substituted into the right-hand side
of the first equation in (\ref{6}) to compute the generator
$\beta(0,\alpha)=\psi(\alpha)= \alpha^2/3\pi$, followed by an
elementary integration of the first of Eqs.(\ref{4}).

Moreover, starting from the two-loop expression $\bar\alpha^{(2)}_{PTh}
(x,;\alpha)$ containing the  $\alpha^2\ell/4\pi^2$ term
we arrive at the second renormalization group approximation (\ref{a2rg})
performing summation of the next-to-leading UV logs.
Comparing solution (\ref{a2rg}) with (\ref{a1rg}) one can conclude that
two-loop correction is extremely essential just in the vicinity of the ghost
pole singularity at $\,x_1=\exp{(3\pi/\alpha)}$. This demonstrates that the
RG method is a regular procedure, within which it is quite easy to estimate
the range of applicability of the results.

  The second order renorm--group solution (\ref{a2rg}) for the invariant
coupling first obtained in~\cite{bs-55b} contains the nontrivial log--of--log
dependence which is now widely known of the two--loop approximation for the
running coupling in quantum chromodynamics (QCD).

  Quite soon~\cite{sh-55} this approach was formulated for the case of QFT
with two coupling constants $g$ and $h$, namely, for a model of pion--nucleon
interactions with self-interaction of pions. To the system of functional
equations for two invariant couplings
\begin{eqnarray}
&&\bar g^2\left(x,y;g^2,h\right)=
\bar g^2\left(\frac x t, \frac y t, \bar g^2(t,y; g^2, h),
\bar h\left(t,y;g^2,h\right)\right)~,\nonumber\\
&&\bar h\left(x,y;g^2,h\right)=
\bar h\left(\frac x t, \frac y t, \bar g^2\left(t,y;g^2,h\right),
\bar h\left(t,y;g^2,h\right)\right)~\nonumber
\end{eqnarray}
there corresponds a coupled system of nonlinear differential equations.
It was analysed \cite{ilya} in one-loop appriximation to carry out the UV
analysis of the renormalizable model of pion-nucleon interaction.

In Refs. \cite{bs-55a,bs-55b,sh-55} and \cite{ilya} the RG was thus directly
connected with practical computations of the UV and IR asymptotics.  Since
then this technique, known as the {\sl renormalization group method} (RGM),
has become the sole means of asymptotic analysis in local QFT.

\subsection{Other early RG applications}

  Another important general theoretical application of the RG method was made
in the summer of 1955 in connection with the (then topical) so-called ghost
pole problem. This effect, first discovered in quantum electrodynamics
\cite{efim,zero2}, was at first thought~\cite{zero2} to indicate a possible
difficulty in QED, and then \cite{dau-bohr,zero3} as a proof of the
inconsistency of the whole local QFT.

  However, the RG analysis of the problem carried out in~\cite{bsh56} on the
basis of massless solution (\ref{2}) demonstrated that no conclusion
obtained with the aid of finite--order computations within perturbation
theory can be regarded as a complete proof.  This corresponds precisely to
the impression, one can get when comparing (\ref{a1rg}) and (\ref{a2rg}). In
the mid 50s this result was very significant, for it restored the reputation
of local QFT. Nevertheless, in the course of the following decade the
applicability of QFT in elementary particle physics remained doubtful in the
eyes of many theoreticians.

In the general case of arbitrary covariant gauge the renormalization
group analysis in QED was carried out in~\cite{tolia}. Here, the point was
that the charge renormalization is connected only with the transverse part of
the photon propagator. Therefore, under nontransverse (for example, Feynman)
gauge the Dyson transformation has a more complex form. This issue has been
resolved by considering the treating the gauge parameter as another coupling
constant.

Ovsyannikov~\cite{oves} found the general solution to the functional RG
equations taking mass into account:
$$
\Phi(y,\alpha)=\Phi\left(y/x, \bar\alpha(x,y;\alpha)\right)~$$
in terms of an arbitrary function $\Phi$ of two arguments, reversible in its
second argument. To solve the equations, he used the differential group
equations represented as linear partial differential equations of the form
(which are now widely known as the Callan---Symanzik equations):
$$
\left\{x\frac{\partial}{\partial x}+y\frac{\partial}{\partial y}
-\beta(y,\alpha)\frac{\partial}{\partial\alpha}\right\}
\bar\alpha(x,y,\alpha)=0~.  $$

The results of this ``period of pioneers'' were collected in the chapter
``Renormalization group'' in the monograph ~\cite{kniga}, the first edition
of which appeared in 1957 (shortly after that translated into English and
French~\cite{book}) and very quickly acquired the status of the ``QFT
folklore''.

\section{Further Bogoliubov's RG Development}

\subsection{Quantum field theory\label{qft}}

  The next decade and a half brought a calm period, during which there was
practically no substantial progress in the renorm--group method.
\vspace{2mm}

 1. New possibilities for applying the RG method were discovered when the
technique of operator expansion at small distances (on the light cone)
appeared \cite{conus}. The idea of this approach stems from the fact that the
RG transform, regarded as a Dyson transformation of the renormalized vertex
function, involves the simultaneous scaling of all its invariant arguments
(normally, the squares of the momenta) of this function. The expansion on the
light cone, so to say, ``separates the arguments", as a result of which it
becomes possible to study the physical UV asymptotic behaviour by means of
the expansion coefficients (when some momenta are fixed on the mass shell).
As an important example we can mention the evolution equations for moments
of QCD structure functions \cite{ap77}.
\vspace{2mm}

 2. In the early 70ies S. Weinberg \cite{steve} proposed the notion of the
{\sl running mass} of a fermion. If considered from the viewpoint of
\cite{tolia}, this idea can be formulated as follows:

{\sl any parameter of the Lagrangian can be treated as a (generalized)
coupling constant, and its effective counterpart should be included into the
renorm-group formalism}.

  However, the results obtained in the framework of this approach turned out
to be, practically, the same as before. For example, the most familiar
expression for the fermion running mass
$$  \bar m(x,\alpha)=m_\mu\left(\frac{\alpha}{\bar\alpha(x,
\alpha)}\right)^\nu~,
$$
in which the leading UV logarithms are summed, was known for the electron
mass in QED (with  $\nu=9/4$) since the mid 50s (see~\cite{lakh}, \cite{bs-55b}).
\vspace{2mm}

 3. The end of the calm period can be marked well enough by the year 1971,
when the renormalization group method was applied in the quantum theory of
non-Abelian gauge fields, in which the famous effect of {\it asymptotic
freedom} has been discovered~\cite{12}.

The one-loop renorm-group expression
$$
\bar\alpha_s^{(1)}(x;\alpha_s)=\frac{\alpha_s}{1+\alpha_s\beta_1\ln x}~, $$
for the QCD effective coupling $\bar\alpha_s$ exhibits a remarkable
UV asymptotic behaviour thanks to $\beta_1$ being positive. This expression
implies, in contrast to Eq.(\ref{a1rg}), that the effective QCD coupling
decreases as $x$ increases and tends to zero in the UV limit. This discovery,
which has become technically possible only because of the RG method use, is
the most important physical result obtained with the aid of the renorm--group
approach in particle physics.
\vspace{2mm}

4. One more interesting application of the RG method in the multicoupling
case, ascending back in 50ies \cite{ilya}, refers to special solutions, so
called separatrices in a phase space of several invariant couplings. These
solutions relate effective couplings and represent a scale invariant
trajectories, like, e.g., $g_i=g_i(g_1)\,$ in the phase space which are
straight lines at the one-loop case.

  Some of them, that are ``attractive" (or stable) in the UV limit, are
related to symmetries that reveal themselves in the high-energy domain.  It
 has been conjectured that these trajectories may be connected to {\it hidden
symmetries of a Lagrangian} and even could serve as a tool to find them. On
this basis the method has been developed \cite{ks76} for finding out these
symmetries. It was shown that in the phase space of the invariant charges the
internal symmetry corresponds to a singular solution that remain
straight-line when taking into account the higher order corrections. Such
solutions corresponding to supersymmetry have been found for some
combinations of Yukawa and quartic interactions.

  Generally, these singular solutions obey the relations
$$
\frac{dg_i}{dt}= \frac{dg_i}{dg_1}\frac{dg_1}{dt}\,,\;\,\, t=\ln x$$
which are known since Zimmermann's paper \cite{Z85} as {\it the reduction
equations}. In the 80ies they have been used \cite{OSZ} (see also review
paper \cite{z-rg} and references therein) in the UV analyzis of
asymptotically free models. Just for these cases the one-loop reduction
relations are adequate to physics.

  Quite recently some other application of this technique has been found in
a supersymmetrical generalizations of Grand Unification scenario in the
Standard Model. It has been shown \cite{EKT,PS,K} that it is possible to
achieve complete UV finiteness of a theory if Yukawa couplings are related to
the gauge ones in a way corresponding to these special solutions,that is to
reduction relations.
\vspace{2mm}

 5. A general method of approximate {\it solution of the massive RG equations}
has been developed \cite{dv81}. Analytic expressions of high level of
accuracy for an effective coupling and one-argument function have been
obtained up to four- and three-loop order \cite{92-3}.
   For example, the two-loop massive expression for the invariant coupling
\begin{eqnarray}
\bar{\alpha}_s(Q^2, m^2)_{\rm rg,2} =\alpha_s\left\{1+\alpha_s A_1(Q^2, m^2)+
\alpha_s\frac{A_2(Q^2,m^2)} {A_1(Q^2, m^2)}\,\ln \left(1+ \alpha_s A_1(Q^2,
m^2) \right)\right\}^{-1}\;
\label{a2rgm}\end{eqnarray}
at small $\,\alpha_s$ values corresponds to adequate perturbation expansion
\begin{eqnarray}
\bar{\alpha}_s(Q^2, m^2)_{\rm pert,2}=\alpha_s\left\{1-\alpha_s A_1(Q^2, m^2)
+ \alpha_s^2A_1^2(Q^2, m^2) - \alpha_s^2\,A_2(Q^2, m^2)+ \dots \right\}\;.
\label{a2pert}\end{eqnarray}
  At the same time, it smoothly interpolates between two massless limits
(with $A_{\ell} \simeq \beta_{\ell} \ln Q^2 +c_{\ell}$) at $\,Q^2 \ll m^2$
and $Q^2\gg m^2\,$ described by equation analogous to Eq.(\ref{a2rg}). In the
latter case it can be represented in the form usual for the QCD practice:
$$
\bar{\alpha}^{-1}_s(Q^2/\Lambda^2)_{\rm rg,2}\to\beta_1\left\{\ln \frac{Q^2}
{\Lambda^2} +b_1 \ln \left(\ln \frac{Q^2}{\Lambda^2} \right) \right\}\,;
\;\; b_1= \frac{\beta_2}{\beta_1^2} \,. $$

 The solution (\ref{a2rgm}) demonstrate, in particular, that the threshold
crossing generally changes the subtraction scheme \cite{mass95}.

   Our investigation \cite{dv81,92-3,92-4} was prompted by the problem of
explicitly taking into account heavy quark masses in QCD. However, the
results obtained are important from a more general point of view for a
discussion of the scheme dependence problem in QFT. The method used could
also be of interest for RG applications in other fields within the situation
with disturbed homogeneity, such as, e.g., intermediate asymptotics in
hydrodynamics, finite-size scaling in critical phenomena and the excluded
volume problem in polymer theory.

  In the paper \cite{92-4} this method was used for the effective couplings
evolution in Standard Model (SM). Here, new analytic solution of a coupled
system of three mass-dependent two-loop RG evolution equations for three SM
invariant gauge couplings has been obtained.
\vspace{2mm}

  6. One more recent QFT development relevant to renorm-group is the {\it
Analytic approach} to perturbative QCD (pQCD). It is based upon the procedure
of {\it Invariant Analytization}\/ \cite{jinr96} ascending to the end of 50ies.

  The approach consists in a combining of two ideas: the RG summation of
leading UV logs with analyticity in the $Q^2$ variable, imposed by spectral
representation of the K\"all\'en--Lehmann type which implements general
properties of local QFT including the Bogoliubov condition of microscopic
causality. This combination was first devised \cite{bls59} to
get rid of the ghost pole in QED about forty years ago. \par

 Here, the pQCD invariant coupling $\bar{\alpha}_s(Q^2)\,$ is transformed
into an ``analytic coupling" $\alpha_{\rm an}(Q^2/\Lambda^2)\equiv {\cal
A}(x)$, which, by constuction, is free of ghost singularities due to
incorporating some nonperturbative structures.

  This analytic coupling ${\cal A}(x)$ has no unphysical singularities in the
complex $Q^2$-plane; its conventional perturbative expansion precisely
coincides with the usual perturbation one for $\bar{\alpha}_s(Q^2)\,$; it has
 no extra parameters; it obeys an universal IR limiting value ${\cal
A}(0)=4\pi/\beta_0\,$ that is independent of the scale parameter $\Lambda$;
it turns out to be remarkably stable with respect to higher loop corrections
and, in turn, to scheme dependence. \par

  Meanwhile, the ``analytized" perturbation expansion \cite{MSS97} for an
observable $F$, in contrast with the usual case, may contain specific
functions ${\cal A}_n(x)$, instead of powers $\left( {\cal A}(x)\right)^n\,$.
In other words, the pertubation series for $F(x)$, due to analyticity
imperative, may change its form \cite{sh99} turning into an asymptotic
expansion \`a la Erd\'elyi over a nonpower set $\{{\cal A}_n(x)\}\,$. \par

\subsection{Ways of the RG expanding}
 As is known, in the early 70ies Wilson~\cite{ken71} succeeded in
transplanting the RG philosophy from relativistic QFT to a quite another
branch of modern theoretical physics, namely, the theory of phase transitions
in spin lattice systems. This new version of the RG was based on Kadanoff's
idea\cite{leo66} of joninig in ``blocks" of few neighbouring spins with
appropriate change (renormalization) of the coupling constant.

  To realize this idea, it is necessary to average spins in each block.
This operation reducing the number of degrees of freedom and simplifying the
system under consideration, preserves all its long-range properties under
a suitable renormalization of the coupling constant. Along with this, the
above procedure gives rise to a new theoretical model of the original
physical system.

   In order that the system obtained by averaging be similar to the original
one, one must also discard those terms of a new effective Hamiltonian which
turns out to be irrelevant in the description of infrared properties. As a
result of this {\it Kadanoff--Wilson decimation}, we arrive at a new model
system characterized by new values of the elementary scale (spacing between
blocks) and coupling constant (of blocks interaction). By iterating this
operation, one can construct a discrete ordered set of models. From the
physical point of view the passage from one model to some other one is an
irreversible approximate procedure. Two passages of that sort applied in
sequence should be equivalent to one, which gives rise to a group structure
in the set of transitions between models. However, in this case the RG is an
approximative and is realized as a semigroup.

   This construction, obviously in no way connected with UV properties, was
much clearer from the general physical point of view and could therefore be
readily understood by many theoreticians. Because of this, in the seventies
the RG concept and its algorithmic structure were successfully carried over
to diverse branches of theoretical physics such as polymer physics \cite{14},
the theory of noncoherent transfer \cite{15}, and so on.

  Apart from constructions analogous to that of Kadanoff--Wilson, in a number
of cases the connection with the original quantum field renorm--group was
established.
This has been done with help of the functional integral representation.
For example, the classic Kolmogorov--type turbulence problem was
connected with the RG approach by the following steps~\cite{16}:
\begin{enumerate}
\item Define the generating functional for correlation functions.
\item Write for this functional the path integral representation.
\item By a change of functional integration variable establish an equivalence
     of the given classical statistical system with some QFT model.
\item Construct the Schwinger--Dyson equations for this equivalent QFT.
\item Use the Feynman diagram technique and perform a finite renormalization.
\item Write down the standard RG equations and use them to find fixed
      point and scaling behavior.
\end{enumerate}
The physics of renormalization transformation in the turbulence problem
is related to a change of UV cutoff in the wave-number variable.
\vspace{2mm}

   Hence, in different branches of physics the RG evolved in two directions:
\begin{itemize}
\item  The construction of a set of models for the physical problem at hand
by direct analogy with the Kadanoff--Wilson approach (by averaging over
certain degrees of freedom) --- in polymer physics, noncoherent transfer
theory, percolation theory, and others;
\item The search for an exact RG symmetry directly or
by proving its equivalence to some QFT: for example, in turbulence
theory ~\cite{16,sasha} and turbulence in plasma~\cite{17}.
\end{itemize}

What is the nature of the symmetry underlying the renormalization group?

a) In QFT the renorm--group symmetry is an exact symmetry of a solution
described in terms of the notions of the equation(s) and some boundary
condition(s). \par
b) In turbulence and some other branches of physics it is a symmetry
  of a solution of an equivalent QFT model.    \par
c) In spin lattice theory, polymer theory, noncoherent transfer theory,
percolation theory, and so on (in which the Kadanoff--Wilson blocking concept
is used) the RG transformation involves transitions inside a set of auxiliary
models (constucted especially for this purpose). To formulate RG, one should
define an ordered set ${\cal M}$ of models $M_i$. The RG transformation
connecting various models has the form $\,R(n)M_i=M_{ni}~$.  Here, the
symmetry can be formulated only in the terms of whole set $\,{\cal M}$. \par

   There is also a purely mathematical difference between the aforesaid RG
realizations. In QFT the RG is a {\it continuous symmetry group}. On the
contrary, in the theory of critical phenomena, polymers, and other cases
(when an averaging operation is necessary) we have an {\it approximate
discrete semigroup}. It must be pointed out that in dynamical chaos theory,
in which RG ideas and terminology can sometimes be applied too, functional
iterations do not constitute a group at all, in general.  An entirely
different terminology is sometimes adopted in the above--mentioned domains of
theoretical physics. Terms like ``the real--space RG", ``the Wilson RG'',
``the Monte--Carlo RG'', or ``the chaos RG'' are in use.

 Nevertheless, the affirmative answer to the question {\it ``Are there
distinct renormalization groups?''} implies no more than what has just been
said about the differences between cases a) and b) on the one hand and c)
on the other.

  For this reason, we shall use notation of the {\sf``Bogoliubov
Renormalization Group"} for the exact Lie group, as it emerged from the QFT
original papers \cite{stp,bs-55a,nc-56} (see also chapter ``Renormalization
Group" in the monograph \cite{kniga,book}) of mid-fifties. This is to make
clear distinction between exact group and the Wilson construction for which
the term ``Renormalization Group" is widely used in the current literature.

\subsection{Functional self--similarity.}

An attempt to analyse the relationship between these formulations on a simple
common basis was undertaken about fifteen years ago~\cite{sh-82}. In this
paper (see also our surveys \cite{sh-84,O-7,KEK}) it was demonstrated that
all the above--mentioned realizations of the RG could be considered in a
unified manner by using only some common notions of mathematical physics.

In the general case it proves convenient to discuss the symmetry underlying
the renorm--group with the aid of a continuous one--parameter transformation
of two variables $x\,$ and $g$
\begin{equation}   \label{10}
R_t:\left\{x\to x^\prime=x/t,~g\to g^\prime=\bar g(t,g)\right\}~.
\end{equation}
Here, $x\,$ is the basic variable subject to a scaling transformation, while
$g\,$ is a physical quantity undergoing a more complicated functional
transformation. To form a group, the transform $R_t$ must satisfy the
composition law
$$
R_tR_\tau=R_{t\tau}~, $$
which yields the functional equation for $\bar g$:
\begin{equation}\label{12}
\bar g(x,g)=\bar g\left(x/t, \bar g(t,g)\right)~.
\end{equation}
This equation has the same form as the functional equation (\ref{3}) for the
effective coupling in QFT in the massless case, that is, at $\,y=0$. It is
therefore clear that the contents of RG equation can be reduced to the group
composition law. \par
  In physical problems the second argument $\,g\,$ of the transformation
usually is related to the boundary value of a solution of the problem under
investigation. This means that the symmetry underlying the RG approach is a
symmetry of a solution (not of equation) describing the physical system at
hand, involving a transformation of the parameters entering the boundary
conditions.

   Therefore, in the simplest case the renorm--group can be defined as a
continuous one--parameter group of transformations of a solution of a
 problem fixed by a boundary condition. The RG transformation affects
the parameters of a boundary condition and corresponds to changing the way in
which this condition is introduced for {\it one and the same solution}.

  Special cases of such transformations have been known for a long time.
If we assume that $F= \bar g\;$ is a factored function of its arguments,
then from Eq.(\ref{12}) it follows that $F(z,f)=fz^k$, with $k$ being
a number. In this particular case the group transform takes the form
$$
P_t~:~\{~z\to z^\prime=z/t~,~~f\to f^\prime=ft^k~\}~,  $$
which is known in mathematical physics long since as a power {\it
self-similarity transformation}. More general case $\,R_t\,$ with functional
transformation law (\ref{10}) can be characterized \cite{sh-82} as a {\it
functional self--similarity} (FSS) transformation.

\subsection{Recent application in mathematical physics\label{mf}}

  We can now answer the question concerning the physical meaning of the
symmetry that underlies FSS and the Bogoliubov renorm--group. As we have
already mentioned, it is not a symmetry of the physical system or the
equations of the problem at hand, but a symmetry of a solution considered as
a function of the essential physical variables and suitable boundary
conditions. A symmetry like that can be related, in particular, to the
invariance of a physical quantity described by this solution with respect to
the way in which the boundary conditions are imposed. The changing of this
way constitutes a group operation in the sense that the group property can be
considered as the transitivity property of such changes.

  Homogeneity is an important feature of the physical systems under
consideration. However, homogeneity can be violated in a discrete manner.
Imagine that such a discrete inhomogeneity is connected with a certain value
of $\,x\,$, say, $\,x=y$. In this case the RG transformation with canonical
parameter $\,t\,$ will have the form:
\begin{equation}
R_t~:~\{~x^\prime=x/t~,~~y^\prime=y/t~,~~g^\prime=\bar g(t,y;g)~\}~.
\end{equation}

The group composition law yields precisely the functional equation (\ref{3}).

The symmetry connected with FSS is a very simple and frequently encountered
property of physical phenomena.  It can easily be ``discovered" in
various problems of theoretical physics: in classical mechanics, transfer
theory, classical hydrodynamics, and so on \cite{O-7,KEK,mamikon,RG91}.

Recently, some interesting attempts have been made to {\it use the RG concept
in classical mathematical physics}, in particular, to study strong nonlinear
regimes and to investigate asymptotic behavior of physical systems described
by nonlinear partial differential equations (DEs).

 About a decade ago the RG ideas were applied by late Veniamin Pustovalov
with co-authors \cite{KP-87-90} to analyze a problem of
generating of higher harmonics in plasma. This problem, after some
simplification, was reduced to a couple of partial DEs with the boundary
parameter -- solution ``characteristic" -- explicitly included. It was proved
that these DEs admit an exact symmetry group, that takes into account
transformations of this boundary parameter, which is related to the amplitude
of the magnetic field at a critical density point. The solution symmetry
obtained was then used to evaluate the efficiency of harmonics generation in
cold and hot plasma. The advantageous use of the RG-approach in solving the
above particular problem gave promise that it may work in other cases and
this was illustrated in \cite{KKP_RG_91} by a series of examples for various
boundary value problems.

 Moreover, in Refs. \cite{O-7,KKP_RG_91} the possibility of devising a
regular method for finding a special class of symmetries of the equations in
mathematical physics, namely, RG-type symmetries, was discussed. The latter
are defined as solution symmetries with respect to transformations involving
parameters that enter into the solutions through the equations as well as
through the boundary conditions in addition to (or even rather than) the
natural variables of the problem present in the equations.

 As it is well known, the aim of modern group analysis \cite{Oves,Ibr}, which
goes back to works of Sophus Lie \cite{Lie}, is to find symmetries of DEs.
This approach does not include a similar problem of studying the symmetries
of solutions of these equations. Beside the main direction of both the
classical and modern analysis, there also remains the study of solution
symmetries with respect to transformations involving not only the variables
present in the equations, but also parameters entering into the solutions
from boundary conditions.

 From the aforesaid it is clear that the symmetries which attracted attention
in the 50s in connection with the discovery of the RG in QFT were those
involving the parameters of the system in the group transformations. It is
natural to refer to these symmetries related to FSS as the {\it RG-type
symmetries}.

  It should be noted that the procedure of revealing the RG symmetry (RGS),
or some group feature, similar to RG regularity, in any partial case (QFT,
spin lattice, polymers, turbulence and so on) up to now is not a regular
one. In practice, it needs some imagination and atypical manipulation
``invented" for every particular case --- see discussion in \cite{umn94}.
By this reason, the possibility to find a regular approach to constructing
RGS is of principal interest.

  Recently a possible scheme of this kind was presented in application to
mathematical model of physical system that is described by DEs. The leading
idea \cite{RG91,KKP_RG_91,Shr_Pisa-95} in this case is based on the fact that
solution symmetry for such system can be found in a regular manner by using
the well-developed methods of modern group analysis. The scheme that
describes devising of RGS is then formulated \cite{KPSh_JMP_98} as follows.

Firstly, a specific RG-manifold should be constructed. Secondly, some
auxiliary symmetry, i.e., the most general symmetry group admitted by this
manifold is to be found. Thirdly, this symmetry should be restricted on
a particular solution to get the RGS. Fourthly, the RGS allows one to improve
an approximate solution or, in some cases, to get an exact solution.

  Depending on both a mathematical model and boundary conditions, the first
step of this procedure can be realized in different ways. In some cases, the
desired RG-manifold is obtained by including parameters, entering into a
solution via equation(s) and boundary condition, in the list of independent
variables. The extension of the space of variables involved in group
transformations, e.g., by taking into account the dependence of coordinates
of renorm--group operator upon differential and/or non-local variables (which
leads to the Lie-B\"acklund and non-local transformation groups \cite{Ibr})
can also be used for constructing the RG-manifold. The use of the
Ambartsumian's invariant embedding method \cite{Ambar} and of differential
constraints sometimes allows reformulations of a boundary condition in a form
of additional DE(s) and enables one to construct the RG-manifold as a
combination of original equations and embedding equations (or differential
constraints) which are compatible with these equations. At last, of
particular interest is the perturbation method of constructing the
RG-manifold which is based on the presence of a small parameter. \par

  The second step, the calculating of a most general group $\cal{G}$ admitted
by the RG-manifold, is a standard procedure in the group analysis and has
been described in detail in many texts and monographs -- see, for example,
\cite{Oves,Sprav,Olver}.

  The symmetry group $\cal{G}$ thus constructed can not as yet be referred to
as a renorm--group. In order to obtain this, the next, third step should be
done which consists in restricting $\cal{G}$ on a solution of a boundary
value problem.  This procedure utilizes the invariance conditionš and
mathematically appears as a ``combining" of different coordinates of group
generators admitted by the RG-manifold.

  The final step, i.e., constructing analytic expression for solution of
boundary value problem on the basis of the RGS, usually
presents no specific problems. A review of results, that were obtained on the
basis of the formulated scheme can be found, for example, in
\cite{KPSh_JMP_98,K_MGA_96,KS_JNOPM_97}.

  Up to now the described regular method is feasible for systems that can be
described by DEs and is based on the formalism of modern group analysis.
However, it seems also possible to extend our approach on physical systems
that are not described just by differential equations. A chance of such
extension is based on recent advances in group analysis of systems of
integro-differential equations~\cite{Mel,KKP_DE_93} that allow
transformations of both dynamical variables and functionals of a solution to
be formulated~\cite{KKP_JNMP_96}. More intriguing is the issue of a
possibility of constructing a regular approach for more complicated systems,
in particular to that ones having an infinite number of degrees of freedom.
The formers can be represented in a compact form by functional (or path)
integrals.

\vspace{3mm}
The author is indebted to D.V. Kazakov, V.F.Kovalev, B.V.Medvedev as well as
to late V.V. Pustovalov for useful discussion and comments.

\end{document}